\documentclass[%
reprint,
amsmath,amssymb,
aps,
prd
]{revtex4-2}

\usepackage{amsmath}%
\usepackage{amssymb}
\usepackage{graphicx}
\usepackage{dcolumn}
\usepackage{bm}

\newcommand{\gcc}{\mbox{g~cm$^{-3}$}}

\newcommand{\B}{\bm{B}}

\newcommand{\vu}{\bm{u}}
\newcommand{\dd}{\mathrm{d}}

\newcommand{\msun}{M_\odot}

\begin{document}

\title[Powerful flares and oscillations of magnetars]{Powerful flares and magneto-elastic oscillations of magnetars}

\author{D. G. Yakovlev}
\email{yak.astro@mail.ioffe.ru}
\affiliation{Ioffe Institute, Politekhnicheskaya street 26, Saint Petersburg, 194021, Russia}%

\date{\today}

\begin{abstract}
Magnetars are neutron stars with superstrong magnetic fields which can exceed $10^{15}$ G. Some magnetars (the so-called soft gamma-repeaters – SGRs) demonstrate occasionally very powerful processes of energy release, which result in exceptionally strong flares of electromagnetic radiation. It is believed that these flares are associated with the presence of superstrong magnetic fields. Despite many hypotheses, the mechanism of these flares remains a mystery. In afterglows of the flares, one has often observed quasi-periodic oscillations (QPOs) of magnetar emission. They are interpreted as stellar vibrations, excited by the flares, which are useful for exploring the nature of magnetar activity. The incompleteness of theories employed to interpret magnetar QPOs is discussed.\\ 
Published in: Zhurnal Experimentalnoi i
Teoreticheskoi Fiziki, Vol. 166 (7) (2024).

\end{abstract}

\maketitle

\section*{1. Introduction}
\label{s:introduc}

Pyotr Leonidovich Kapitza, to whom this issue of Journal of Experimental and Theoretical Physics (ZhETF) is dedicated, made an outstanding contribution to studies of very strong magnetic fields \cite{1933Kapitza}. Perhaps he would have liked magnetars -- the natural  
laboratories of superstrong magnetic fields.

Neutron stars are the most compact of all stars. They are well known  
astrophysical objects, but are still fascinating because of extreme physical conditions in and around them. They 
contain superdense matter with superstrong magnetic fields
in the presence of enormous gravitational forces. Many properties of neutron 
stars (for example, the equation of state and the composition of matter in inner layers)
are still poorly understood.

\begin{figure}
	\centering
	\includegraphics[width=0.4\textwidth]{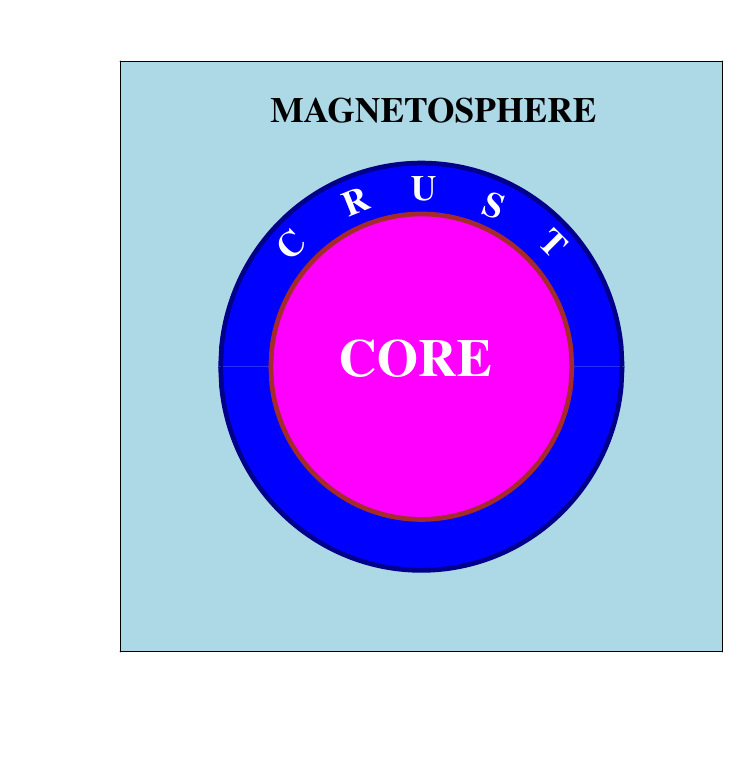}%
	\caption{
		Schematic structure of a neutron star.
		A massive and bulky core of superdense nuclear matter is surrounded by an outer shell (crust) containing an elastic crystal of atomic nuclei. A magnetar possesses superstrong  magnetic fields and is surrounded by a powerful magnetosphere.
	}
	\label{f1}
\end{figure}

A schematic structure of a neutron star is shown in Fig.\ \ref{f1}. One can distinguish two
main internal layers (e.g., \cite{HPY2007}): the outer shell, often called the crust, and the inner core.
At a typical neutron star mass,
$M \sim 1.4\,\msun$ ($\msun$ is the solar mass) its radius is $R \sim 12$ km.
The crust consists mostly of ions (atomic nuclei), electrons, and
(at densities $\rho \gtrsim 4 \times 10^{11}$ $\gcc$) free neutrons. It is $\sim 1$ km thick, has a mass of $\sim 0.01\, \msun$. The density
at the crust bottom is about half the standard nuclear density $\rho_0$, with 
$\rho_0 \approx 2.8 \times 10^{14} \, \gcc$. The atomic nuclei in the crust form usually Coulomb crystals. Beneath the crust, there is a massive and bulky core, containing liquid nuclear matter; its composition and equation of state
are not reliably known. The central density of the star reaches several
$\rho_0$. 

This paper is devoted to magnetars (as reviewed, e.g. in
Ref.\ \cite{2017KasB}) which are neutron stars with extraordinary strong magnetic fields. Some of them 
form as a special class of sources called soft gamma-ray repeaters (SGRs). Occasionally, SGRs demonstrate huge energy releases (up to $\sim 10^{46}$ erg), observed as powerful flares of electromagnetic radiation, which then fade. It is thought that these processes are driven by superstrong magnetic fields. There are many models 
(e.g., Ref.\ \cite{2017KasB}) but the nature of magnetar flares is still unknown, and it will not be discussed here. 	

It is important, that the flares are accompanied by observed quasi-periodic oscillations (QPOs) of the magnetar emission 
at certain frequencies. These are assumed to be the frequencies of stellar oscillations excited by the flares.   
In principle, a correct interpretation of the observed QPOs can provide useful 
information on the parameters of magnetars, on the
strength and geometry of their magnetic fields, and on the mechanism of their flaring activity. This motivates studies of the QPO problem.

The existence of QPOs in magnetar flares was theoretically predicted by Duncan {\cite{1998Duncan} in 1998.
The first QPOs were discovered after observations of the giant flare of SGR 1900+14 
(27 Aug.\ 1998) and the hyper-flare of SGR 1806--20 (27 Dec.\ 2004). This was done by  
careful processing observational data in the 2005--2006 \cite{2005Israel,2005Strohmayer,2006Watts}, which initiated serious studies of QPOs. These observations, as well as observations of other flares of
magnetars, have been processed and reprocessed many times.
(e.g., \cite{2011Hambaryan,2014Huppen,2014Huppenkothen,2018Pumpe}). The data on the SGR 1806--20 hyperflare seem most representative, apparently due to the exceptionally huge energy release in the event.
	
The observed frequencies  $\nu$ of magnetar QPOs fall in a wide range from 
tens of Hz to several kHz. The QPOs are usually divided
into low-frequency ($\nu \lesssim 150$ Hz) and  high-frequency ones (at higher $\nu$).  
The detection of QPOs in magnetar flares has given rise to a variety of calculations  
and interpretations of oscillation frequencies (e.g,
\cite{2006Levin,
		2006Glampedakis,
		2007Levin,
		2007Sotani,
		2008Sotani,
		2008Lee,
		2009CD,
		2009SE,
		2009Colaiuda,
		2011Colaiuda,
		2011vanHoven,	
		2011Gabler,
		2012Colaiuda,
		2012vanHoven,
		2012Sotani,
		2012Gabler,
		2013Sotani,
		2013Gabler,
		2013Gabler1,
		2014Passamon,
		2016Gabler,
		2016Link,
		2018Gabler,
		2018Sotani,
		2023Yak1} and references therein). 
		
This paper is a logical continuation of the previous
work (Refs.\ \cite{2009SE} and \cite{2023Yak1}). It provides new arguments to prove that 
previous calculations of magnetar QPOs 
have mostly dealt with an incomplete set of solutions. The paper is arranged as follows. Firstly,
the formalism is briefly described in Sec.\ 2. Then
oscillation modes and torsional oscillations of non-magnetic are  are outlined (in Secs.\ 3 and 4, respectively).
In Sec.\ 5 we discuss magneto-elastic oscillations assuming
that magnetic field effect is
sufficiently weak. In Sec.\ 6 this case is studied for
a pure dipole magnetic field in the stellar crust, and the consideration is extrapolated to higher magnetic fields.
In Sec.\ 7, the possibility of applying the results for
interpreting the observed QPOs is discussed. In Sec.\ 8 the results are summarized and unsolved problems are formulated.

\section*{2. Formalism}
\label{s:formalism}

Oscillations of magnetars are described by the standard formalism of magneto-elastic oscillations of neutron stars. The formalism is well known (e.g., \cite{2012vanHoven}); it sufficient to outline the basic points. 
For simplicity, the equations are presented neglecting  relativistic effects. These effects will be included
in Sec.\ 5. Magneto-elastic 
oscillations are
mediated by elastic forces 
of crystal lattice in the stellar crust and by elastic deformations 
(Alfv\'en perturbations) of magnetic field lines everywhere where the 
field is present. 
	
The star is assumed to have a stationary magnetar field $\B(\bm{r})$ 
($\sim 10^{14}-10^{16}$ G), which is not strong enough to cause a noticeable distortion 
of stellar shape from the spherical one. The oscillation equations are obtained by linearizing the equations of motion of magnetized matter assuming  
the field is frozen into the matter. The unperturbed configuration of the star is thought to be
spherically symmetric. Under these conditions, it
is sufficient to study the  
oscillations of incompressible matter 
where matter elements
move only along spherical surfaces. Then the perturbations of pressure and density
are absent, and the emission of gravitational waves is suppressed.  The perturbations 
excite small velocities of matter elements ${\bf v}(\bm{r},t)$, small
displacements of these elements ${\bf u}(\bm{r},t)$, and small variations of magnetic magnetic field
${\bf B}_1(\bm{r},t)$. All these variations oscillate in time as $\exp({\rm i}\omega t)$, where $\omega= 2 \pi \nu$ is the angular oscillation frequency,
and $\nu$ is the cyclic frequency.
This overall oscillating factor in the equations can be dropped, leading to the 
stationary wave equation for small (complex) amplitudes $\vu(\bm{r})$ and ${\B_1}(\bm{r})$, and for the 
oscillation frequency $\omega$: 
\begin{equation}
	\rho \omega^2 \vu= {\bf T}_\mu + {\bf T}_{B}.
	\label{e:newton}     
\end{equation}
Here ${\bf T}_\mu$ and ${\bf T}_{ B}$ are the volume densities of forces
(with minus sign) determined,
respectively, by the crystal elasticity
and magnetic field stresses.
In the first case
\begin{equation}
	{\bf T}_{\mu i}=- \frac{\partial \sigma_{ik}}{\partial x_k},\quad
	\sigma_{ik}= \mu \,\left(\frac{\partial u_i}{\partial x_k} 
	+\frac{\partial u_k}{\partial x_i} \right), 
	\label{e:Tsigma}        
\end{equation}
where is $\sigma_{ik}$ is the tensor of shear deformations and $\mu$ is the
shear viscosity (in the isotropic crystal approximation). In the case
of magnetic forces one has
\begin{equation}
	{\bf T}_{ B}= \frac{1}{4 \pi}\, \B {\bf \times }
	{\rm curl}\, \B_1, \quad \B_1={\rm curl}(\vu {\bf \times} \B).
	\label{e:TB}    
\end{equation}
	
These equations should be supplemented with boundary conditions. Since the crystal
exists only in the stellar crust, the radial components of viscous stresses
must vanish at the outer and inner boundaries of the crust. The conditions for the magnetic
field depend on the formulation of the problem. Alfv\'en perturbations can
propagate into the core and magnetosphere of the star.

\section*{3. General remarks}
\label{s:comments}
	
Let us begin with a few remarks. 
It is well known that the shear modulus $\mu$ determines characteristic
propagation speed $v_\mu$ of elastic shear deformations  in a crystal. As follows from calculations
(e.g., \cite{2020KY}),
these deformations are mainly located near the bottom of the  crust, at $\rho \sim 10^{14}$ \gcc. Then, under typical conditions, one 
comes to the estimate
\begin{equation}
	v_\mu \sim \sqrt{\mu/ \rho} \sim 10^8~~~{\rm cm~s^{-1}}.
	\label{e:vmu}   
\end{equation}
	
As for magnetic perturbations, they propagate with the 
Alfv\'en velocity
$v_{\rm A}$ which, for the same conditions, can be estimated as 
\begin{equation}
	v_{\rm A} = \frac{B}{\sqrt{4 \pi \rho}} \sim 3 \times 10^7\,B_{15}~~~{\rm cm~s^{-1}},
	\label{e:vA}   
\end{equation}
where $B_{15}$ is the magnetic field in units of $10^{15}$ G. The velocity $v_{\rm A}$ can noticeably decrease within the star and
increase toward the surface.

The velocities (\ref{e:vmu}) and (\ref{e:vA}) 
become close at 
\begin{equation}
	B \sim B_{\mu} \sim 3 \times 10^{15}~~{\rm G}.
\end{equation}
This characteristic field strength 
reveals the existence of three
regimes of magneto-elastic oscillations (Table \ref{tab1}).
		
\begin{table}
\caption{Three regimes of magneto-elastic
oscillations of magnetars.\label{tab1}}
\begin{tabular}{ccc}
			\hline
			\hline
			Regime	& Condition	& Leading mechanism  \\ 
			\hline
			I & $B \ll B_{\mu}$  &  shar waves in crystalline matter	  
			\\
			II & $B \sim B_\mu$ & shear and Alfv\'en waves     
			\\
			III & $B \gg B_{\mu}$  &  Alfv\'en waves 
			\\	  
			\hline
\end{tabular}
\end{table}
	
In regime I, the oscillations are mainly regulated by 
elastic shear waves in the stellar crust; Alfv\'en perturbations are
driven by these elastic shear stresses and weakly affect the oscillation frequencies. Such oscillations are almost completely localized in the crust being determined by the microphysics of matter and by the properties 
of $\bm{B}(\bm{r})$ in the crust.

The efficiencies of shear and Alfv\'en waves in regime II are comparable. 
Alfv\'en perturbations can propagate beyond the crust
(e.g. \cite{2006Levin,2006Glampedakis}) and spread
over the whole star. Their calculation requires knowledge of the entire microphysics of the magnetar, which contains many uncertainties, including the equation of state, superfluidity and superconductivity of the stellar core, as well as magnetic field configuration within it.
	
Finally, in regime III, the oscillations are mainly regulated by Alfve\'n waves. The elasticity of crystal becomes nearly or even
fully negligible (e.g., \cite{2008Sotani,2009CD}). 
	
This classification of oscillations is schematic. In particular, it does not take into account
possible forbidden frequency intervals of Alfve\'n oscillations in the stellar core
(e.g., \cite{2012vanHoven}) in which case 
the oscillations can be locked in the crust
even at very strong magnetic fields. The effects of penetration of Alfv\'en perturbations from the crust to the core and back can also be important. They can lead to frequency variations, damping, and loss of coherence of  crustal oscillations.
	
The employed approximation of incompressible  magneto-elastic oscillations (Sec.\ 2) also deserves a comment.  
Types of neutron star oscillations are numerous (e.g., \cite{1988McDermott} and references therein). Magneto-elastic oscillations are suitable because
their frequencies are low enough to explain magnetar QPOs. The shear and Alfv\'en velocities, $v_\mu$ and $v_{\rm A}$, are generally 
much lower than the speed of ordinary sound, that is determined
by the full pressure of dense 
stellar matter. Frequencies of oscillations
of many types are higher than magneto-elastic ones.
	
\section*{4. Elastic oscillations of non-magnetic
crust}
\label{s:B=0}
	
Such oscillations are often called torsional. They are basic for  studying  magneto-elastic oscillations.
Their theory began in the 1980s in the classical works of Hansen and Cioffi \cite{1980Hansen},
Schumaker and Thorne \cite{1983ST}, 
and McDermott et al.\ \cite{1988McDermott} long before the discovery of magnetar QPOs. After
the discovery, the interest to
the theory was renewed (e.g., \cite{2007Samuel,2009Andersson,2012Sotani,2013Sotani,2013aSotani,2016Sotani,2017Sotani,2017aSotani,2018Sotani,2019Sotani} and references therein).
	
Any torsion oscillation
mode is characterized  
by three quantum numbers: (1) $n=0,1,2,\ldots$ is the number of radial nodes of the wave function, (2) the orbital number
$\ell$, which in this problem runs the values $\ell=2,3,\ldots$, (3) the azimuthal number $m$ which takes 
integer values from $-\ell$ to $\ell$.
	
In the spherical coordinates ($r,\theta,\phi$), a stationary wave function $\vu(\bm{r})$ has 
only two non-trivial components: $\vu_\phi$
and $\vu_\theta$ (since $\vu_r=0$). 
These can be written as
(e.g., \cite{2023Yak1})
\begin{eqnarray}
	u_\phi (r,\theta,\phi) & = &	r Y(r)\, 
	{\rm e}^{{\rm i}m \phi}\, 
	\frac{\dd\,P_{\ell}^m}{\dd \theta},
	\label{e:uphi} \\
	u_\theta (r,\theta,\phi) & = &	r Y(r)\, {\rm e}^{{\rm i}m \phi}\,
	\frac{{\rm i}m P_{\ell}^m}{\sin \theta},
	\label{e:utheta}
\end{eqnarray}	
where $P_\ell^m(\cos \theta)$ is an associated Legendre polynomial,
and $Y(r)=Y_{n \ell}(r)$ is a dimensionless radial wave-function 
satisfying the equation
\begin{equation}
	Y''+\left( \frac{4}{r}+ \frac{\mu'}{\mu} \right)	Y'+
	\left[ \frac{\rho}{\mu}\,\omega^2
	-\frac{(\ell+2)(\ell-1)}{r^2}   \right]
	Y=0 .
	\label{e:Y}	
\end{equation}
Prime means derivative with respect to $r$. 
These oscillations are localized in the crystalline crust,
$r_1 \leq r \leq r_2$, where $r_1$ is the radius
of the crust-core interface, and $r_2$ is the outer radius of the crystallization zone which is very close
to the radius of the star. At both boundaries, radial elastic stresses should vanish, $Y'(r_1)=Y'(r_2)=0$.
The frequencies of torsional oscillations are degenerate in $m$:
$\omega=\omega_{\mu n\ell}$ (the index  $\mu$ indicates 
the elastic shear nature of these
oscillations); the functions $Y(r)$ are independent of $m$.  The value $Y_0=Y(r_2)$ characterises the angular amplitude of oscillations (in radians) at the outer edge of the crystallization region.
If $m=0$, crustal matter oscillates only along parallels 
($\vu_\theta=0$), but at $m \neq 0$ there
appear
meridional motions. The value of $m$ strongly influences
the geometry of displacements $\vu(\bm{r})$ and the angular dependence of the energy density of oscillations.
A specific stellar model affects only $Y(r)$, while angular dependences of $\vu(\bm{r})$ stay standard.
	
Torsional oscillations of neutron stars are divided into fundamental 
($n=0$) and ordinary ($n>0$) ones. For the fundamental oscillations, a very good approximation  
is the weak deformability of the crystal,
in which case $Y$ is almost independent of $r$ (e.g., \cite{2020KY}). In this case, $\omega_{\mu 0 \ell } \approx \tfrac{1}{2} \omega_{\mu 0} \sqrt{(\ell+2)(\ell-1)}$, where 
$\omega_{\mu 0}$ is the basic frequency (at $\ell=2$), the lowest for all torsion oscillations.

The frequencies of ordinary torsion oscillations ($n>0$) are higher and strongly increase with increasing $n$. At a fixed $n$, there is a bunch 
of close frequencies which grow weakly
with increasing $\ell$ (showing "fine  splitting" with respect to $\ell$).
Corresponding 
wave functions $Y_{n \ell}(r)$ depend on $\ell$ rather weakly (e.g., \cite{2023Yak}). Since torsional oscillation frequencies do not depend 
on $m$, one usually sets $m=0$ for
finding the oscillation spectrum, without using the  
states with $m \neq 0$.
	
Torsional oscillations may carry a lot of energy. 
For example, let us choose a neutron star model 
with a nucleonic core and 
the modern BSk21 equation of state of dense matter (described, e.g., in Ref.\ \cite{BSk2013}). For a $1.4\, \msun$
neutron star, the stellar radius 
is $R=12.6$ km and the crust-core radius is  $r_1=11.55$
km. According to the results of Ref.\ \cite{2023Yak}, the oscillation energy of the basic mode ($n=0$, $\ell=2$,
$\nu_{\mu 0}=23.0$ Hz)
is $E_{\rm vib} \approx 10^{49}\, Y_0^2$ 
erg. At a swing angle  $\approx 0.1^\circ$ 
($Y_0 \sim 1.7 \times 10^{-3}$ rad)
of oscillations at the outer edge of the crystalline crust,
we get $E_{\rm vib} \sim 3 \times 10^{43}$ erg. In this case,
shear stresses in a vibrating crust are still far from the crystal-breaking
limit \cite{2020KY}.

\section*{5. Oscillations dominated by elasticity of the crust}
\label{s:regime1}
	
This is the simplest regime I of magneto-elastic oscillations 
(Table \ref{tab1}).
In this case magnetic fields are not too high
($B \ll B_\mu$) and can be taken into account by perturbation theory, considering the wave functions of pure torsional oscillations (Sec.\  4) as zero-order wave functions, and the quantity ${\bf T}_{B}$ in Eq.\  (\ref{e:newton}) as a small perturbation. In numerous studies of magneto-elastic oscillations 
(e.g.,
\cite{2006Levin,
		2006Glampedakis,
		2007Levin,
		2007Sotani,
		2008Lee,
		2009Colaiuda,
		2011vanHoven,
		2011Colaiuda,
		2012Colaiuda,
		2012vanHoven,
		2011Gabler,
		2012Gabler,
		2013Gabler,
		2013Gabler1,
		2016Gabler,
		2018Gabler}), 
the states with
$m \neq 0$ have been 
ignored. In this way one dealt with
incomplete spectrum
of magneto-elastic
oscillations.

The exceptions were the paper by Shaisultanov and Eichler
{\cite{2009SE} and the recent paper {\cite{2023Yak1}. The authors of Ref.\ \cite{2009SE} argued that the  magnetic field removes  
the degeneracy of torsion frequencies. In a magnetic field, these frequencies should split into a series of frequencies, which can be treated as the Zeeman effect in magnetars. The effect was correctly described and evaluated, but the work did not attract much attention. Ref.\ \cite{2023Yak1} was devoted to  developing these ideas. It proposed a simple algorithm for calculating the oscillation frequencies in the first-order  perturbation theory for a wide class of $\B$-fields in the magnetar crust. 
For illustration, the Zeeman splitting of fundamental oscillations ($n=0$) in a dipole crustal magnetic field at $2 \leq \ell \leq 5$ was calculated.
			
Here we extend the consideration  
of magneto-elastic oscillations in the first-order perturbation theory for the fundamental modes ($n=0$). The details of the
theory were presented in
{\cite{2023Yak1}. Here we
mention them only briefly.
In the formulated approach, it is sufficient
to assume that the oscillations are localized in the elastic crust. As in Ref.\ \cite{2023Yak1}, we assume
that the crustal magnetic field is axially symmetric about the magnetic axis: only the field components $B_r(r,\theta)$ and $B_\theta(r,\theta)$ are different from zero. In this case
\begin{equation}
	\omega_{\ell m}^2 = \omega_{\mu  \ell}^2 + \omega_{B \ell m}^2,
	\label{e:omegaB2}
\end{equation}
where $\omega_{\mu \ell}$ is the frequency of purely torsional
oscillations (Sec.\ 4), and $\omega_{B \ell m}$ is the small `magnetic' correction; $\ell$ and $m$ have the same meaning as
in the wave functions of zero-order 
approximation; see Eqs.\ (\ref{e:uphi}) and (\ref{e:utheta}).
				
The expressions for $\omega_{\mu \ell}$ and $\omega_{B \ell m}$ are given in \cite{2023Yak1}. 
In Sec.\ 6 of Ref.\ \cite{2023Yak} it is also described which changes should be 
introduced 
to the theory to account for relativistic effects. According to \cite{2023Yak1},
\begin{equation}
				\omega_{\mu \ell}^2= 
				\frac{(1-x_{\rm g*})\int_{\rm crust} \dd V\, \mu }{\int_{\rm crust} \dd V \, (\rho + {\rm P}/c^2)\, r^2},
				\label{e:freqtorsn=0}    
\end{equation}
\begin{equation}
				\omega_{B \ell m}^2=
				\frac{	(1- x_{\rm g*})\, \frac{1}{4 \pi} \int_{\rm crust} \dd V \, I_B}
				{\Xi(\ell,m)\,
					\int_{\rm crust} \dd V \, 
					(\rho+{\rm P}/c^2)\, r^2 } .
				\label{e:omegaB5}	
\end{equation}
Here P is the pressure of dense matter; $c$ is the speed of light;
$\dd V=r^2\, \dd r\,
\sin \theta \, \dd \theta \, \dd \phi$ is the volume element in the approximation of locally flat crust; integration is over crystalline matter. The factor $(1-x_{\rm g*})$ 
approximates the gravitational redshift of a squared oscillation frequency 
for a distant observer,
$x_{\rm g*}=2GM_*/(c^2r_*)$, $G$ is the gravitational constant.
Furthermore, $r_*$ is the radius 
of any point in the crust (results being almost independent of its particular choice \cite{2023Yak}),
$M_*$ is the gravitational mass inside a sphere
of radius $r_*$. The quantity
\begin{equation}
				\Xi(\ell,m)= \frac{2 \ell (\ell+1) (\ell+m)!}{(2 \ell+1)(\ell-m)!}
\label{e:Xi}
\end{equation}
is a convenient normalization factor, and $I_B$ is a combination of $B_r$, $B_\theta$, $P_\ell^m$ and their derivatives (see  Eq.\ (18) in \cite{2023Yak1}) quadratic in the magnetic field, making $\omega^2_{B\ell m} \propto B^2$. In this case, the oscillations are located in the crust and do not depend on the $\B(\bm{r})$ configuration 
outside the crust.

\section*{6. Case of the dipole field}
\label{s:dipole}
				
By way of illustration, following Ref.\ \cite{2023Yak1},  we  consider a purely dipole 
magnetic field in the stellar crust. Then $B_r=B_0\, \cos \theta \, (R/r)^3$ and
$B_\theta=\tfrac{1}{2}\,B_0\, \sin \theta \, (R/r)^3$. Here $B_0$ is the 
field strength at the magnetic pole on the stellar surface. The  field 
removes degeneracy of frequencies $\omega_{\mu \ell}$, but only partly: according to (\ref{e:omegaB2}) the frequency
$\omega_{\mu \ell}$ splits into a series of $\ell+1$ components $\omega_{\ell m}$, where $m=0,1,\dots \ell$. The frequency $\omega_{\ell 0}$ appears non-degenerate, 
while the frequencies with $m>0$ remain degenerate twice (correspond 
to $\pm m$ states). The Zeeman splitting is determined by  $\omega_{B \ell m}$ given by Eq.\ (\ref{e:omegaB5}). For the pure dipole field
\begin{eqnarray}
				&&I_B = - \frac{B_0^2}{4} \left(\frac{R}{r}\right)^6 \left[  
				P'^2 (1+ 3 \cot^2 \theta)
				-3 P'P'' \cot \theta + P' \right.
				\nonumber \\
				&& \left.  \times P''' +
				\frac{m^2}{\sin^2 \theta}
				\left(-P'^2-10\, P P' \cot \theta
				+ 8P^2 \cot^2 \theta   \right)         \right].
				\label{e:IBdipole}	
\end{eqnarray}
Here $P=P_\ell^m(\cos \theta)$, prime denotes differentiation with
respect to $\theta$.
Then
\begin{equation}
				\omega_{B \ell m}^2
				= \frac{{B_0^2 r_2^3 }\, \LARGE[ \left( {r_2/r_1}  \right)^3 -1  
					\LARGE]}{12 \pi \int_{r_1}^{r_2} \dd r \, r^4 (\rho+{\rm P}/c^2)}\, \zeta_{\ell m}, 
				\label{e:omegaBdipole}
\end{equation}
where
\begin{equation}
				\zeta_{\ell m}=
				\frac{1}{B_0^2 \, \Xi(\ell,m)}\,
				\int_0^\pi \sin \theta \, \dd \theta \,I_B(R,\theta). 
				\label{e:zeta}
\end{equation}
Our Eqs.\ (\ref{e:IBdipole}) and (\ref{e:omegaBdipole}) correspond to Eqs.\ (26) and (27) in Ref.\ \cite{2023Yak1}. The latter contain typos, which are corrected here. All calculations in Ref.\ \cite{2023Yak1} were performed using  correct formulae.
				
In Ref.\ \cite{2023Yak1} the factors $\zeta_{\ell m}$  were calculated and approximated by 
\begin{equation}
				\zeta_{\ell m}=c_{0}(\ell)+c_2(\ell)m^2
				\label{e:fit}	
\end{equation}
at $\ell \leq 5$; the values of
$c_0(\ell)$ and $c_2(\ell)$ were tabulated. 
Now the values of $\zeta_{\ell m}$ have been calculated 
up to $\ell=15$.  
Corresponding values of $c_0(\ell)$ and $c_2(\ell)$ can be 
fitted as
\begin{equation}
				c_0(\ell)=0.721\,[(\ell-2)(\ell+1)]^{0.954},
				\label{e:c0}
\end{equation}
\begin{equation}
				c_2(\ell)=\frac{2}{3}-\frac{0.766\,(\ell-2)^{1.09}}{
					1+0.532\,(\ell-2)^{1.15}}.
				\label{e:c2}
\end{equation}
The fit accuracy is a few per cent, which seems quite satisfactory. Equations 
(\ref{e:fit})--(\ref{e:c2}) are valid for a dipole magnetic field in the crust; fields of other configurations will be studied separately.
				
\begin{figure}
\centering
\includegraphics[width=0.45\textwidth]{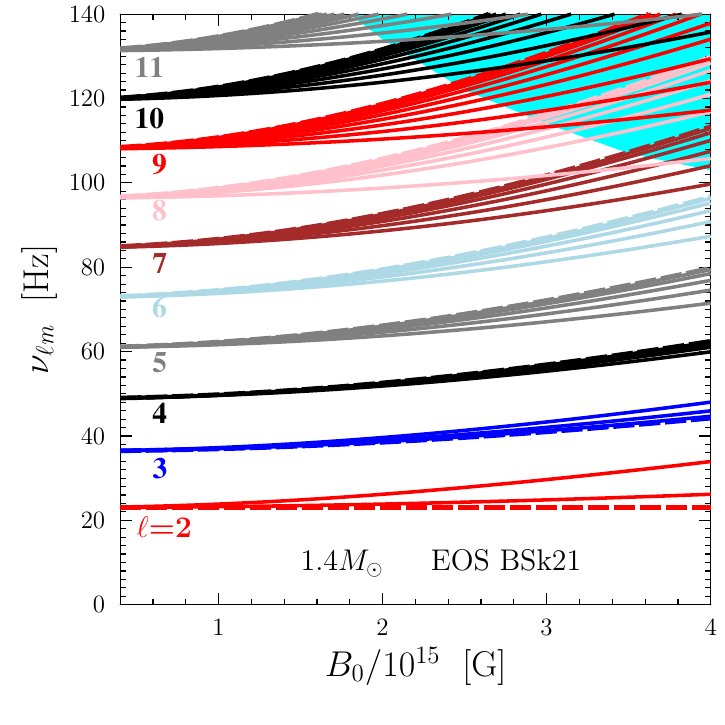}%
\caption{
		The frequencies of magneto-elastic oscillations 
		of a $1.4 \, \msun$ neutron star 
		versus the 
		field strength $B_0$ at the magnetic pole on the stellar surface. 
		Each series of frequencies corresponds to a fixed  $\ell=2,
		\ldots 11$ and contains a bunch of $\ell+1$ 
		Zeeman components. The components with 
		$m=0$ are shown by dashed lines.  
		The region which contains many quasi-crossings of Zeeman  components is darkened (as detailed in the text).  
					}
\label{f2}
\end{figure}	 
				
The results for dipole fields are
illustrated below, extending thus the consideration of Ref.\ \cite{2023Yak1} to a wider frequency range.
				
Figure \ref{f2} shows the dependence of oscillation frequencies on $B_0$ at $n=0$, $\ell=2,\ldots 11$ and different $m$. As in Ref.\ \cite{2023Yak1}, we consider
the $1.4\,\msun$ star with the Bsk21 equation of state, mentioned in Sec.\ 4. Calculations are performed using 
Eqs.\ (\ref{e:omegaB2}), 
(\ref{e:freqtorsn=0}) and (\ref{e:omegaBdipole}). Figure  \ref{f2} is similar to Fig.\ 1b from Ref.\ \cite{2023Yak1}, but covers wider frequency range $\nu \leq 140$ Hz (instead of 80 Hz in \cite{2023Yak1}).

According to the results of Sec.\ 3, Fig.\ \ref{f2}
shows two regimes of magneto-elastic oscillations: regime
I of field strengths much smaller than $B_\mu \sim 3 \times 10^{15}$ G, and regime II of intermediate field strengths (Table \ref{tab1}). The equations employed  
are strictly valid only in regime I. In the figure, they are extrapolated to the intermediate regime. The possibility of such extrapolation requires confirmation (see below).
				
At $B_0 \leq 4 \times 10^{14}$ G and $\nu < 140$ Hz, Fig.\ \ref{f2} shows 10 frequencies of fundamental torsional oscillations 
(Sec.\ 4) which  
are actually unaffected by the magnetic field. However, as $B_0$ grows up, each of these frequencies splits noticeably  
into Zeeman components: 10 initial frequencies decompose into 75 branches.

At not too high $B_0$, one can clearly see
10 separate bunches of curves corresponding to certain $\ell$.
The oscillation frequencies 
in each bunch differ by the values of
$m$. 
In agreement with the results of \cite{2023Yak1}, the branches of oscillations with $m=0$ (dashed lines)
at $\ell=2$ and 3 lie below other branches in a bunch, while at higher $\ell$ they become higher than the others (this inversion is possibly specific for the dipole magnetic field). The higher  $\ell$, the richer the splitting, and the smaller the value of $B_0$ at which this splitting starts to be visible.

At $\ell>3$, the lowest branch of oscillations in any bunch corresponds to the highest $m=\ell+1$. Interestingly, as $\ell$ increases, such curves become more horizontal and depend weaker on $B_0$. In other words,
at high $m$ the frequencies $\nu_{\ell m}(B_0)$ 
approach the frequencies of torsional oscillations
$\nu_{\mu \ell}$ of an non-magnetic star (Sec.\ 4).

Starting from $B_0 \gtrsim 1.5 \times 10^{15}$ G 
and $\ell \sim 11$, in the upper right corner of 
Fig.\ \ref{f2}, there appears a special region of frequencies and magnetic fields  in which the magneto-elastic oscillations behave in a complicated way. If $B_0$ increases to
the highest depicted values ($4 \times 10^{15}$ G), this region descends to  
 $\nu \sim 90$ Hz (and at higher $B_0$ it
will descend further). In this region the two effects, neglected in the calculations, can be especially important.

Firstly,  
with increasing $B_0$, the oscillation modes from different bunches begin to show quasi-crossings (Fig.\  \ref{f2}). The identity of individual bunch is lost, and the region becomes densely filled with allowed oscillation frequencies. The behavior of curves near quasi-crossing points requires further analysis. As usual, in the vicinity of these points, the oscillations of converging  
modes interact with each other, and their frequencies
are distorted. 
				
Secondly, in the presence of pronounced damping and loss of coherence of crustal oscillations due to the transfer of vibrational energy by Alfve\'n waves into the stellar core, the oscillation modes can 
acquire finite shifts and widths (this effect is  
expected to be amplified with increasing $B_0$ and $\nu$). The frequencies of crustal oscillations are thus capable of blurring and shifting. However, there can exist forbidden 
frequency intervals (e.g., \cite{2012vanHoven} and references therein), which may prevent 
penetration of Alfve\'nic waves into the core.
				
It is clear that both effects are mutually related and require self-consistent consideration.
It is impossible to correctly calculate quasi-crossings 
without a reliable theory of interaction between crustal oscillations and Alfv\'en perturbations in the core. Great efforts have been spent on the construction of such a theory
(e.g,
				\cite{2006Levin,
					2006Glampedakis,
					2007Levin,
					2008Sotani,
					2009Colaiuda,
					2011vanHoven,
					2011Colaiuda,
					2012Colaiuda,
					2012vanHoven,
					2011Gabler,
					2012Gabler,
					2013Gabler,
					2013Gabler1,
					2016Gabler,
					2018Gabler})
but only for axially symmetric perturbations ($m=0$). 
There is no theory at $m \neq 0$, it
is a difficult task for future studies. It seems that both effects are most important in the special region of high frequencies and fields, while at lower $\nu$ and $B_0$ they are weaker.

\section*{7. Discussion}
\label{s:discuss} 
				
\begin{figure}
\centering
	\includegraphics[width=0.45\textwidth]{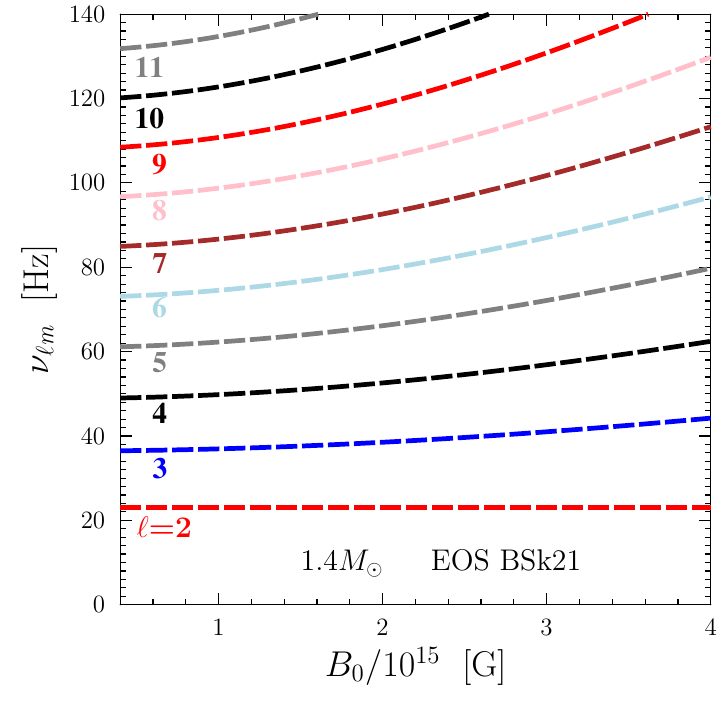}%
	\caption{Same as in Fig.\ \ref{f2}, but we left 
	only the oscillations with $m=0$, which were  considered
	in the majority of publications.
					}
\label{f3}
\end{figure}

All the above effects can be important for interpreting observed frequencies of magnetar 
QPOs. Below we extend the interpretation of QPOs with
 complete theoretical spectrum (it was started
in Ref.\ \cite{2023Yak1}) using the data from   
the hyperflare of SGR 1806--20 and the giant
flare of SGR 1900+14.
Now we add a few higher-frequency QPOs.  
The unsolved problems of quasi-crossing of magneto-elastic oscillation frequencies and the interaction of crustal oscillations with Alfve\'n oscillations in the stellar core are not considered here. Therefore, as in Ref.\ \cite{2023Yak1}, our consideration is illustrative and may
be particularly inaccurate at sufficiently high $\nu$ and $B_0$.

The available observational data on  magnetar QPOs have
been analyzed many times.
The results are summarized, for example, in Ref.\ \cite{2018Gabler}. They have been widely used by many authors and will be used below. An exception is Ref.\ \cite{2018Pumpe}, whose authors expressed doubt in significance of measured low-frequency QPOs based on Bayesian model-independent extraction of noise; their conclusion requires further confirmation.
				
It has already been noted that in many interpretations the oscillation modes with $m \neq 0$ were ignored.
For illustration, Fig.\ \ref{f3} presents only the
frequencies of $m=0$ oscillation modes, instead of 
all of them  
in Fig.\ \ref{f2}. Of the 75 modes shown
in Fig.\ \ref{f2}, only 10 are left. Clearly,
they constitute an  incomplete set
of theoretical curves. Using such a set, an interpretation of observations can be questionable. In
particular, all quasi-crossings of oscillation modes 
in Fig.\ \ref{f3} disappear.
				
The reason for neglecting solutions with $m \neq 0$ is that the displacements $\vu(\bm{r})$ of oscillating stellar matter for such solutions depend not only on $r$ and $\theta$, but also on the angle $\phi$; see, e.g., Eqs.\ (\ref{e:uphi}) and (\ref{e:utheta}). 
In other words, for axially symmetric
 magnetic fields $\B(r,\theta)$,
considered by most researchers, the perturbed quantities $\vu$ and $\B_1$ at
$m \neq 0$ turn out to be axially asymmetric. However, the axial symmetry of perturbations was usually postulated leading to the loss of solutions with $m \neq 0$.
				
Let us add that,
as seen from Fig.\ \ref{f3}, the frequency $\nu_{20}$ (the lowest dashed line) does not depend on $B$ at all (see also Ref.\ \cite{2023Yak1}). This result is valid in the first-order perturbation theory for weak dipole crustal magnetic fields (Sec.\
5). In reality, it means that an  expansion of 
$\nu_{20}(B_0)$ in powers of $B_0^2$ 
should  not contain the $B_0^2$-term. Higher-order terms
can be present, although their calculation requires much effort. 
But according to Ref.\ \cite{2007Sotani}, devoted to oscillations of a neutron star crust
with a dipole field, the 
$B_0^2$-term is not zero. This paradox was resolved by 
noting \cite{2012vanHoven}, that the solution in \cite{2007Sotani} had been sought by expanding  
$u_\phi$ into a series of functions (\ref{e:uphi}) with 
different $\ell$ at $m=0$. The sum over $\ell$ in \cite{2007Sotani}
was artificially truncated, that  
was actually equivalent to solving an exact
non-dipole magnetic field problem. This does not
prove the  
$B_0^2$-term is nonzero.
				
\begin{figure}
	\centering
	\includegraphics[width=0.45\textwidth]{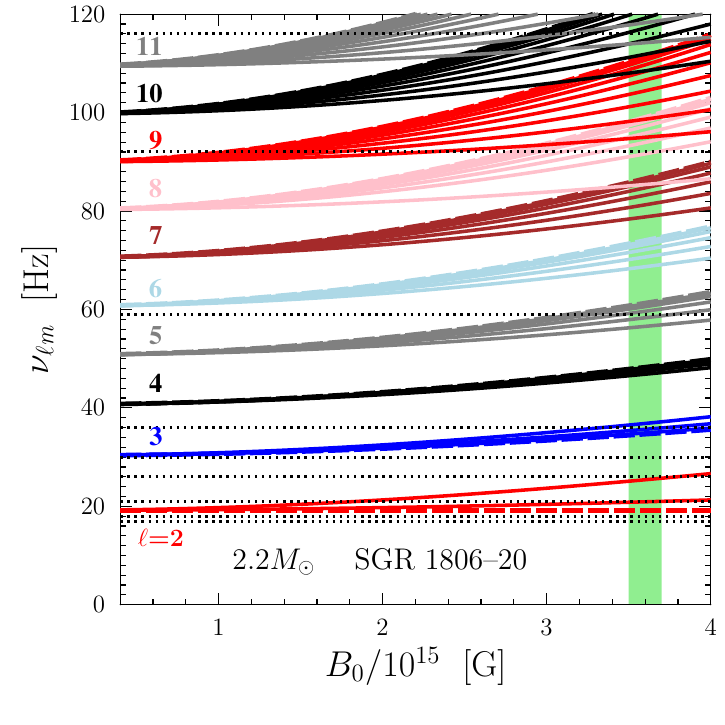}%
	\caption{
	Same as in Fig.\ \ref{f2}, but for a  $2.2\, \msun$ star
	compared to QPO frequencies (dotted horizontal lines) observed from the hyper-flare of 
	SGR 1806--20. The vertical green 
	band shows a  possible
	range of $B_0$, simultaneously
	consistent with some 
	observed QPOs (as detailed in the text).
	}
	\label{f4}
\end{figure}
				
The largest number of QPOs was detected by processing observations of the hyper-flare of SGR 1806--20. The low-frequency QPOs, which are discussed below, were detected at  
18, 26, 30, and 150 Hz, and (with lower confidence) at 17, 21, 36, and 59 Hz. 
				
Can these QPOs be interpreted as fundamental magneto-elastic oscillations of a single star (with the same mass, radius, and internal structure) possessing the same crustal dipole field? This question was raised in Ref.\ \cite{2023Yak1}, where theoretical calculations were limited to $\ell \leq 5$ and could explain QPO frequencies $\nu \leq 60$ Hz. It turned out that for a 
$M=1.4\, \msun$ star, only three of seven such frequencies (17, 18, 21, 26, 30, 36, and 59 Hz) could be interpreted:
26, 30, and 59 Hz, assuming $B_0 \approx (3.2-3.4)
\times 10^{15}$ G (Fig.\ 2a in \cite{2023Yak1}). The lowest-frequency QPOs could not be explained in this way. However, leaving the same equation of state of the neutron star matter (BSk21), but increasing the stellar mass to
$2.2 \, \msun$ (with the $2.27 \, \msun$ maximum mass limit), one could slightly lower all theoretical frequencies due to a stronger gravitational redshift of  oscillation frequencies for a more massive (and compact) star. In this case (Fig.\ 2b in \cite{2023Yak1}), it was possible to explain six frequencies except for
one (30 Hz), assuming $B_0 \approx 
(3.5 - 3.7) \times 10^{15}$~G.
				
By adding new calculations of this paper (up to $\ell = 11$), it is possible to explain
all but one (30 Hz) of the observed low-frequency QPOs of the  SGR 1806--20 hyperflare assuming the same field $B_0 \approx 
(3.5-3.7) \times 10^{15}$ G as in the \cite{2023Yak1}. 
This is shown in Fig.\ \ref{f4}, that is similar to Fig.\ 2b in \cite{2023Yak1} but extended now to 120 Hz.  An explanation for the highest selected QPO frequency
(150 Hz) is not shown to simplify the figure, but it is evident 
because of 
very densely spaced theoretical oscillation branches at $\nu> 90$ Hz. We do not worry on the failure to explain 
the 30 Hz QPO \cite{2023Yak1}: no attempt has been made to seriously explain the observations. 
The 
$M=2.2\, \msun$ model was chosen as an example and was not varied. The required magnetic
field $B_0$ corresponds to regime II (Sec.\ 3), where quantitative accuracy of theoretical frequencies can be questioned. In addition, a purely dipole magnetic field has been assumed, whereas possible deviations from pure dipole can
change theoretical results. In any case, the proposed complete theoretical set of frequencies of magneto-elastic oscillations greatly simplifies  theoretical interpretation of low-frequency magnetar QPOs.

\begin{figure}
\centering
\includegraphics[width=0.45\textwidth]{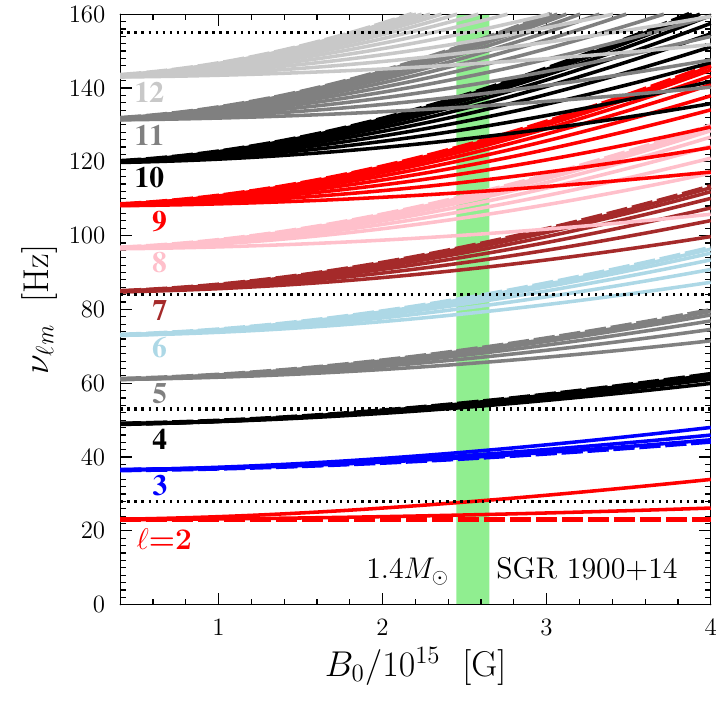}%
					\caption{
	Same as in Figs. \ref{f2} and \ref{f4}, 
	compared with the QPO frequencies (dotted horizontal lines) observed from the giant flare of 
	SGR 1900+14.
}
\label{f5}
\end{figure}
				
Moreover, in Ref.\ \cite{2023Yak1} we tried to interpret the QPOs observed in the giant flare of SGR 1900+14. Four low-frequency QPOs (28, 53, 84, and 155 Hz) were detected. The two lowest frequencies were easily
explained by the $1.4 \, \msun$ neutron star model (Fig.\ 3 in \cite{2023Yak1}). By taking the same model and now increasing the theoretical frequencies to 160 Hz, it is possible to explain  
(Fig.\ \ref{f5}) all  
four QPOs with $B_0 \approx (2.42-2.62) \times 10^{15}$ G. Just as for  
Fig. \ref{f4}, 
a more serious interpretation seems premature.
				
Let us emphasize that $B \sim B_\mu$ has often been treated as a valid estimate of magnetar fields for various reasons (e.g., \cite{2018Gabler}).

\section*{8. Conclusions}
\label{s:conclude}
				
We have attempted to develop the theory of magneto-elastic oscillations of magnetars. 
We have employed the standard assumption that these oscillations are excited during flares of magnetars and are observed as  quasi-periodic oscillations (QPOs) at the decay phases 
of the flares (Sec.\ 1 and references therein). Correct interpretation of observations can provide useful information about the parameters of magnetars, their magnetic fields, and the nature of their flares.
				
Following the results of Refs.\ \cite{2009SE,2023Yak1}, the completeness of theoretical QPO models has been studied, especially because many previous works
neglected Zeeman splitting of magnetar oscillations. We have used a simple model of low-frequency magneto-elastic oscillations without nodes of radial wave function in the magnetar crust 
assuming a purely dipole crustal magnetic field. We have shown  that neglecting the Zeeman effect leads to essentially incomplete set of oscillation modes.
In the case of axially symmetric $\B(\bm{r})$, this simplification
reproduces only the oscillations accompanied by axially symmetric vibrational displacements of matter elements $\vu(\bm{r})$ 
and magnetic field $\B_1(\bm{r})$. In this way it
misses a wide range of
oscillations modes in which the displacements $\vu(\bm{r})$ and $\B_1(\bm{r})$ are axially asymmetric.
We have demonstrated that the full set of oscillations gives a qualitatively different 
oscillation spectrum and can significantly change theoretical interpretation of observed
QPOs.
								
Therefore, the construction of complete set of magneto-elastic oscillations has only begun. Serious efforts are needed to complete it. Here are some of the problems.
								
Even in the most reliable regime I of relatively low magnetic fields, only low-frequency oscillations have been considered, 
without nodes of wave function along the radius. Generalization to the case of oscillations with nodes
($n>0$) can be done without difficulty. Instead of the dipole field, 
it is easy to study poloidal magnetic fields of other
types. It is also easy to consider the case in which a toroidal magnetic field is also present. 
In addition, our results
are obtained in the approximation of a locally flat stellar crust (e.g., \cite{2023Yak,2023Yak1}). It would be useful to solve the problem in full General Relativity.
This would be especially important for the cases in which Alfv\'en perturbations
propagate outside the crust.

The oscillation regime I, which has been studied rather reliably, is insufficient for interpreting the observations.
It seems that the intermediate regime II ($B \sim B_\mu$) is more  important for this purpose. Such oscillations can penetrate into the core of the star, which makes the above consideration quantitatively
inaccurate (although it may be qualitatively applicable, especially
for lowest frequencies). A firm study of oscillations
in this regime is more complicated because the calculations should  include 
microphysics of the stellar core for many possible models (superfluidity and superconductivity in the core, different magnetic field configurations there, etc.). There is also 
an important problem of interaction of Alfv\'en oscillations in the core with crustal oscillations. The energy of crustal vibrations can flow into the core, which
can lead to damping and loss of coherence of the crustal oscillations. All these effects
have been studied for oscillations induced by axially symmetric perturbations. The important case of axially asymmetric perturbations has not been explored.
				
Another area of research is to improve the microphysics of neutron star matter that affects magnetar oscillations. In particular, one can improve calculations of the shear modulus in the crust, consideration of the delicate effects of superfluidity and superconductivity of crustal matter, nuclear interactions, nuclear pasta effects 
at the bottom of the crust, etc. (see, e.g., \cite{2009Andersson,2012Sotani,2013Sotani,2013aSotani,2016Sotani,2017Sotani,2017aSotani,2018Sotani,2019Sotani,2023Yak,2023Yak1,2023Zemlyak}
).   
				
Finally, it is appropriate to list the main original results of this work.
The studies of low-frequency magneto-elastic oscillations in Ref.\ \cite{2023Yak1}
are extended to higher frequencies (Sec.\ 6, Fig. \ref{f2}). Simple approximations are
derived for the coefficients $c_0(\ell)$ and 
$c_2(\ell)$, Eqs.\ (\ref{e:c0}) and (\ref{e:c2}). They allow one to calculate the
oscillation spectrum at $\nu \lesssim 150$ Hz. 
Quasi-crossings of oscillation frequencies with different
$\ell$ are pointed out in the darkened region in Fig.\ \ref{f2}, where allowable frequencies are densely spaced, but in fact can quickly decay through interaction of crustal and Alfv\'en oscillations of the core.
Following Ref.\ \cite{2023Yak1}, possible interpretations  
of 
low-frequency QPOs,
observed in the hyperflare of SGR 1806--20 (Fig.\ \ref{f4}) and the giant flare of SGR 1900+14 (Fig.\ \ref{f5}), are further discussed. 
The necessity  for joint consideration of quasi-crossings of modes and interactions
of crustal oscillations with Alfv\'en perturbations in the stellar core is stressed. It is stated that,
despite considerable efforts of many authors, the theory of magneto-elastic oscillations is far from being completed.

This work was performed within the Work Program (number FFUG-2024-0002) of A.~F.\ Ioffe Institute.
The author is grateful to M.~E.~Gusakov, E.~M.~Kantor, and A.~I.~Chugunov for comments and critical remarks to the previous paper \cite{2023Yak1}, which were useful for writing this one.


				\newcommand{\araa}{Ann. Rev. Astron. Astrophys.}
				\newcommand{\aap}{Astron. Astrophys.}
				\newcommand{\aj}{Astron. J.}
				\newcommand{\apjl}{Astrophys. J. Lett.}
				\newcommand{\apjs}{Astrophys. J. Suppl. Ser.}
				\newcommand{\apss}{Astrophys. Space Sci.}
				\newcommand{\mnras}{Mon. Not. R. Astron. Soc.}
				\newcommand{\pasa}{Publ. Astron. Soc. Aust.}
				\newcommand{\pasj}{Publ. Astron. Soc. Jpn.}
				\newcommand{\pasp}{Publ. Astron. Soc. Pac.}
				\newcommand{\sovast}{Sov. Astron.}
				\newcommand{\ssr}{Space Sci. Rev.}

{}


\begin{thebibliography}{}\label{sec:TeXbooks}					
					
					\bibitem{1933Kapitza}	
					P.~L. Kapitza, {\it Experimental research in strong
						magnetic fields},
					Physics -- Uspekhi {\bf 36 (4)}, 288 (1993).
					
					\bibitem{ST1983}
					S.~L.~{Shapiro}, A.~A.~{Teukolsky},
					{\it {Black holes, white dwarfs, and neutron stars: The physics of
							compact objects}}, Wiley-Interscience, New York  (1983).
					
					\bibitem{HPY2007}
					P.~{Haensel}, A.~Y.~{Potekhin}, D.~G.~{Yakovlev},
					{\it {Neutron Stars. 1. Equation of State and Structure}}, Springer,
					New York (2007).
					
					
					\bibitem{2017KasB}
					V.~M.~{Kaspi}, A.~M.~{Beloborodov}, 
					{Annual Rev. Astron. Astrophys.}  {\bf 55},~261 (2017).
					
					
					\bibitem{1998Duncan}
					R.~C. {Duncan}, 
					{ \apjl}  {\bf 498},~L45 (1998).
					
					
					\bibitem{2005Israel}
					G.~L.~{Israel}, T.~{Belloni}, L.~{Stella}, Y.~{Rephaeli}, D.~E.~{Gruber}, P.~
					{Casella}, S.~{Dall'Osso}, N.~{Rea}, M.~{Persic}, R.~E.~{Rothschild}, 
					{\apjl} {\bf 628},~L53 (2005).
					
					
					\bibitem{2005Strohmayer}
					A.~L.~{Watts}, T.~E.~{Strohmayer}, 
					\apjl\ {\bf 632}, L111 (2005). 
					
					\bibitem{2006Watts}
					A.~L.~{Watts}, T.~{Strohmayer}, 
					{\apjl} {\bf 637},~L117 (2006).
					
					
					\bibitem{2011Hambaryan}
					V.~{Hambaryan}, R.~{Neuh{\"a}user},~K.~D. {Kokkotas}, 
					{\aap} {\bf 528},~A45 (2011).
					
					\bibitem{2014Huppen}
					D.~{Huppenkothen}, L.~M.~{Heil}, A.~L. {Watts},
					E.~{G{\"o}{\u{g}}{\"u}{\textcommabelow s}}, 
					{ \apj}  {\bf 795},~114 (2014).
					
					
					\bibitem{2014Huppenkothen}
					D.~{Huppenkothen}, C. {D'Angelo}, A.~L. {Watts}, L. {Heil}, M. {van der Klis},
					A.~J. {van der Horst}, C. {Kouveliotou}, M.~G. {Baring}, E.
					{G{\"o}{\u{g}}{\"u}{\textcommabelow s}}, J. {Granot}, Y. {Kaneko}, L.
					{Lin}, A. {von Kienlin}, G. {Younes}, 
					{\apj}  {\bf 787},~128 (2014).
					
					\bibitem{2018Pumpe}
					D. {Pumpe}, M. {Gabler}, T. {Steininger}, T.~A. {En{\ss}lin}, 
					{\aap} {\bf 610},~A61 (2018).
					
					
					\bibitem{2006Levin}
					Y. {Levin}, 
					{\mnras} {\bf 368},~L35 (2006).
					
					
					\bibitem{2006Glampedakis}
					K. {Glampedakis}, L. {Samuelsson}, N. {Andersson},
					{\mnras} {\bf 371},~L74 (2006).
					
					
					\bibitem{2007Levin}
					Y. {Levin}, 
					{ \mnras} {\bf 377},~159 (2007).
					
					\bibitem{2007Sotani}
					H. {Sotani}, K.~D. {Kokkotas}, N. {Stergioulas},
					{\mnras}  {\bf 375},~261 (2007).
					
					\bibitem{2008Sotani}
					H. {Sotani}, K.~D. {Kokkotas}, N. {Stergioulas},
					{\mnras} {\bf 385}, L5 (2008).
					
					\bibitem{2008Lee}
					U. {Lee}, 
					{\mnras} {\bf 385}, 2069~(2008).
					
					
					\bibitem{2009CD}
					P.~{Cerd{\'a}-Dur{\'a}n}, N. {Stergioulas}, J.~A. {Font}, 
					{ \mnras} {\bf 397},~1607 (2009).
					
					
					\bibitem{2009SE}
					R. {Shaisultanov}, D. {Eichler},
					{\apjl} {\bf 702},~L23 (2009).
					
					
					\bibitem{2009Colaiuda}
					A. {Colaiuda}, H. {Beyer}, K.~D. {Kokkotas}, 
					{ \mnras}  {\bf 396},~1441 (2009).
					
					\bibitem{2011Colaiuda}
					A.~{Colaiuda}, K.~D. {Kokkotas}, 
					{\mnras}  {\bf 414},~3014 (2011).
					
					
					
					
					\bibitem{2011vanHoven}
					M. {van Hoven}, Y. {Levin}, 
					{\mnras} {\bf 410},~1036 (2011).
					
					\bibitem{2011Gabler}
					M. {Gabler}, P. {Cerd{\'a}-Dur{\'a}n}, J.~A. {Font}, E. {M{\"u}ller}, N.
					{Stergioulas}, 
					{\mnras}  {\bf 410},~L37 (2011).
					
					\bibitem{2012Colaiuda}
					A. {Colaiuda}, K.~D. {Kokkotas}, 
					{\mnras}  {\bf 423},~811 (2012).
					
					
					
					\bibitem{2012vanHoven}
					M. {van Hoven}, Y. {Levin}, 
					{\mnras}  {\bf 420}, 3035 (2012).
					
					\bibitem{2012Sotani}
					H.~{Sotani}, K.~{Nakazato}, K.~{Iida}, K.~{Oyamatsu},
					{\prl} {\bf 108},~201101 (2012).
					
					\bibitem{2012Gabler}
					M. {Gabler}, P. {Cerd{\'a}-Dur{\'a}n}, N. {Stergioulas}, J.~A. {Font}, E.
					{M{\"u}ller}, 
					{\mnras} {\bf 421},~2054 (2012).
					
					\bibitem{2013Sotani}
					H.~{Sotani}, K.~{Nakazato}, K.~{Iida}, K~{Oyamatsu}, 
					{\mnras} {\bf 434},~2060 (2013).
					
					
					\bibitem{2013Gabler}
					M. {Gabler}, P. {Cerd{\'a}-Dur{\'a}n}, J.~A. {Font}, E. {M{\"u}ller}, N.
					{Stergioulas}, 
					{ \mnras} {\bf 430},~1811 (2013).
					
					\bibitem{2013Gabler1}
					M.~{Gabler}, P. {Cerd{\'a}-Dur{\'a}n}, N. {Stergioulas}, J.~A. {Font}, E.
					{M{\"u}ller}, 
					{\prl} {\bf 111},~211102 (2013).
					
					\bibitem{2014Passamon}
					A. {Passamonti}, S.~K. {Lander}, 
					{\mnras}  {\bf 438},~156 (2014).
					
					
					\bibitem{2016Gabler}
					M.~{Gabler}, P. {Cerd{\'a}-Dur{\'a}n}, N. {Stergioulas}, J.~A. {Font}, E.
					{M{\"u}ller}, 
					{\mnras} {\bf 460},~4242 (2016).
					
					\bibitem{2016Link}
					B. {Link}, C.~A. {van Eysden},
					{ \apjl}  {\bf 823},~L1 (2016).
					
					
					\bibitem{2018Gabler}
					M. {Gabler}, P. {Cerd{\'a}-Dur{\'a}n}, N., {Stergioulas}, J.~A. {Font}, E.
					{M{\"u}ller},
					{\mnras}  {\bf 476},~4199 (2018).
					
					
					\bibitem{2018Sotani}
					H.~{Sotani}, K.~{Iida}, K.~{Oyamatsu},
					{\mnras} {\bf 479},~4735 (2018).
					
					\bibitem{2023Yak1}
					D.~G.~Yakovlev,
					Universe {\bf 9 (12)}, 504 (2023).
					
					
					\bibitem{2020KY}
					A.~A.~{Kozhberov}, D.~G.~{Yakovlev}, 
					{ \mnras} {\bf 498},~5149 (2020).
					
					
					
					\bibitem{1980Hansen}
					C.~J. {Hansen}, D.~F. {Cioffi}, 
					{\apj}  {\bf 238},~740 (1980).
					
					\bibitem{1983ST}
					B.~L.~{Schumaker}, K.~S.~{Thorne}, 
					{\mnras} {\bf 203},~457 (1983).
					
					\bibitem{1988McDermott}
					P.~N.~{McDermott}, H.~M.~{van Horn}, C.~J. {Hansen},
					{ \apj} {\bf 325},~725 (1988).
					
					\bibitem{2007Samuel}
					L.~{Samuelsson}, N.~{Andersson}, 
					{\mnras} {\bf 374},~256 (2007).
					
					
					\bibitem{2009Andersson}
					N.~{Andersson}, K.~{Glampedakis}, L.~{Samuelsson},
					{\mnras} {\bf  396},~894 (2009).
					
					
					
					
					
					
					
					\bibitem{2013aSotani}
					H.~{Sotani}, K.~{Nakazato}, K~{Iida}, K.~{Oyamatsu}, 
					{\mnras} {\bf 428},~L21 (2013).
					
					
					\bibitem{2016Sotani}
					H.~{Sotani},
					{\prd} {\bf 93},~044059 (2016).
					
					
					\bibitem{2017aSotani}
					H.~{Sotani}, K.~{Iida}, K.~{Oyamatsu}, 
					{\mnras} {\bf  464},~3101 (2017).
					
					
					\bibitem{2017Sotani}
					H.~{Sotani}, K.~{Iida}, K.~{Oyamatsu},
					{\mnras} {\bf 470},~4397 (2017).
					
					
					
					\bibitem{2019Sotani}
					H.~{Sotani}, K.~{Iida}, K.~{Oyamatsu}, 
					{\mnras} {\bf 489},~3022 (2019).
					
					
					
					
					
					
					
					
					
					
					
					
					\bibitem{2023Yak}
					D. G. {Yakovlev}, 
					{\mnras} {\bf 518},~1148 (2023).
					
					
					
					
					\bibitem{BSk2013}
					A.~Y. {Potekhin}, A.~F. {Fantina}, N. {Chamel}, J.~M. {Pearson}, S. {Goriely}, 
					{Astron. Astrophys.} {\bf 560},~A48
					(2013).
					
					\bibitem{2023Zemlyak}
					N.~A. {Zemlyakov}, A.~I. {Chugunov}, 
					Universe {\bf 9 (5)}, 220 (2023).
					
					
					
					
					
\end{thebibliography}
\end{document}